\documentclass[journal,12pt,draftclsnofoot,onecolumn]{IEEEtran}
\makeatletter

\newcommand{\Rmnum}[1]{\expandafter\@slowromancap\romannumeral #1@}
\makeatother

\usepackage{epsfig}
\usepackage{amsopn}
\usepackage{subfigure}
\usepackage{cite}
\ifCLASSINFOpdf

\else
\fi
\usepackage[cmex10]{amsmath}

\usepackage{algorithmic}
\usepackage{algorithm}
\begin{document}
\title{Access Strategy in Super WiFi Network Powered by Solar Energy Harvesting: A POMDP Method}

\author{\authorblockN{Tingwu~Wang, Jian~Wang, Chunxiao Jiang, Jingjing Wang and
	Yong~Ren}
	\small\authorblockA{
		Department of Electronic Engineering, Tsinghua University, Beijing, 100084, P. R. China\\
        E-mail: wtw12@mails.tsinghua.edu.cn, chinaeephd@gmail.com, \{jian-wang, jchx, reny\}@tsinghua.edu.cn}}
\maketitle

\begin{abstract}
The recently announced Super Wi-Fi Network proposal in United States is aiming to enable Internet access in a nation-wide area.
As traditional cable-connected power supply system becomes impractical or costly for a wide range wireless network,
new infrastructure deployment for Super Wi-Fi is required.
The fast developing Energy Harvesting (EH) techniques receive global
attentions for their potential of solving the above power supply problem.
It is a critical issue, from the user's perspective, how to make efficient network selection and access strategies. Unlike traditional wireless networks, the battery charge state and tendency in EH based networks have to be taken into account when making network selection and access, which has not been well investigated.
In this paper, we propose a practical and efficient framework for multiple base stations access strategy in an EH powered Super Wi-Fi network.
We consider the access strategy from the user's perspective,
who exploits downlink transmission opportunities from one base station.
To formulate the problem, we used Partially Observable Markov Decision Process (POMDP) to model users' observations on the base stations' battery situation and decisions on the base station selection and access. Simulation results show that our methods are efficacious and significantly outperform
the traditional widely used CSMA method.
\end{abstract}
\IEEEpeerreviewmaketitle
\section{Introduction}
In order to expand the coverage area of wireless network,
many algorithms and implementations have been proposed.
Recently, the Federal Communications Commission published the Super Wi-Fi proposal,
aiming to make use of lower-frequency white spaces between television channel frequencies
and create a nationwide wireless network.
However, the ambitious task of building a countrywide network is confronted with many obstacles.
An inevitable problem is how to deploy practical backhaul and power supply system.
Specifically, traditional cable-based systems may be not appropriate, considering the cost of deploying and maintaining the network.
Despite all the above difficulties, many successful experimental deployments of Super Wi-Fi system are accomplished accordingly.
Wireless backhual has been proven to be effective as a replacement of cable backhaul \cite{30}.
Meanwhile, the fast developing energy harvesting (EH) technology provides an ideal power supply mode,
which could make use of a wide range of ambient energy including piezoelectric, thermal, solar energy, etc.\\
\indent Given that the deployment of EH network is just emerging,
new wireless protocols and modification are required, as some preliminary studies pointed out in \cite{27}. Specifically, from users' perspective, when confronted with EH powered network, how to make efficient network selection and access strategies is a practical and important problem. Different from traditional networks, a prominent issue in EH powered networks is that the energy state and tendency of one base station (BS) has to be taken into account when making BS selection and access strategies.
In the literature, the access strategy problem in traditional wireless networks has been studied extensively, among which
Markov Decision Process was used,
with some challenges and solutions summarized in \cite{23}.
Access process among different users was deemed as a typical game process,
and thus game theory, summarized in \cite{22}
as well as pricing theory investigated in \cite{7}, can be applied, respectively.
In \cite{5}, the access strategy towards multiple base stations with negative externality was considered.
Recently, a POMDP MAC layer opportunistic access was proposed by Dr. Zhao in \cite{zhao1}, and
a learning based approach to access between packet bursts was studied in \cite{kae1}.
\\
\indent However, all the aforementioned studies on users' access strategy did not consider the energy harvesting situation.
In this paper, we propose a user access strategy for the fast booming Super Wi-Fi network, focusing on the influence of BS's battery states.
As long time transmission is not guaranteed in EH network,
the leaving and arriving of users are more frequent.
Instead of using a static system model,
we consider a model where both the number of accessing users and the charge of the BS's battery are dynamic.
Moreover, a new stochastic process for formulating users' arrival and departure processes, as well as a battery state transition based on a quasi-static formulation
are combined to describe the system state transition.
Considering that in EH powered wireless network, where the full knowledge of the system is unrealistic,
we build a Partially Observable Markov Decision Process (POMDP) model to formulate the access strategy problem, i.e., users make the network access strategy by using the partially observed BSs' battery states' information.
It is worth to mention that although our work focuses on solar energy harvesting,
the conclusion and the algorithm could be generalized to any access problems in EH powered network.
\\
\indent The rest of this paper is organized as follows.
We describe the system model in Section \Rmnum{2}.
In Section \Rmnum{3}, the POMDP access strategy is presented.
Furthermore, we explain how to formulate the POMDP states model in order to obtain the optimal access strategy.
Also a suboptimal strategy is proposed for simplifying the calculation.
In Section \Rmnum{4}, we evaluate the performance of our proposed approach in contrast to several famous traditional algorithms, followed by our conclusion in Section \Rmnum{5}.
\section{System Model}
As shown in Fig. 1, in an EH network with multiple BSs, where each BS is supplied by EH devices and connected to the server by wireless backhaul.
In each time slot, the BS harvests a certain quantity of solar energy,
denoted as \(E_H\), and stores it into its battery.
At the same time, to serve the users connected to the BS, energy \(E_T\) is consumed for transmission.
We denote the battery quantity of the BS as \(Q_B\),
which can be any value from \(0\) to the battery volume \(B_M\). In every times slot, the BSs could serve multiple users,
the number of which is denoted as \(S_U = i,\, i = \,0,\,1,\,\ldots,\,N_U-1\), where \(N_U\) is the maximum number of users the BS could serve simultaneously due to limited spectrum and coding ability. Note that different BSs in the network could have different battery volume \(B_M\) and maximum serving users \(N_U\).

\begin{figure}\label{fig1}
\centering
\includegraphics[width=7.5cm]{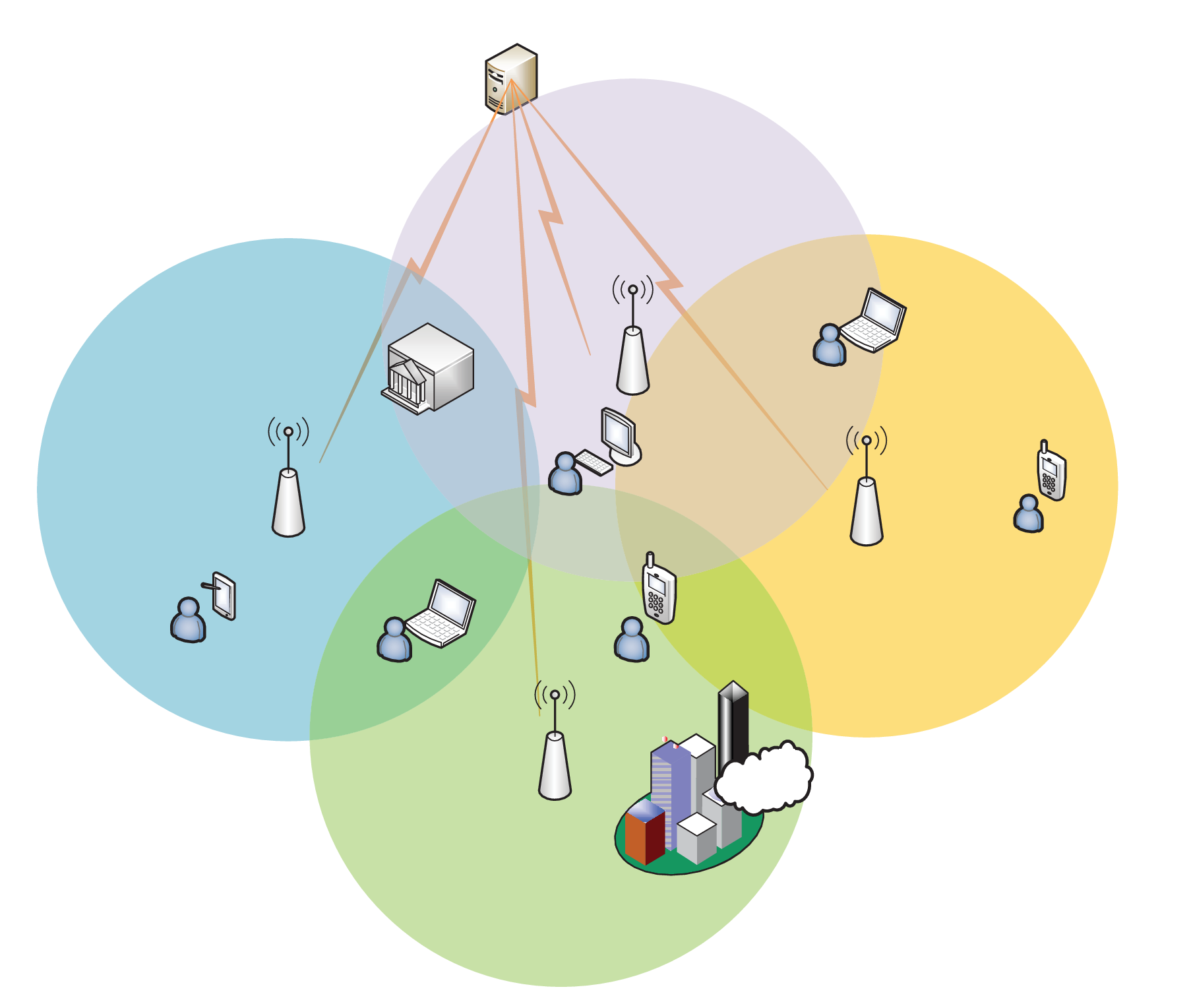}
\caption{A schematic map of Super Wi-Fi system.}
\end{figure}
At the start of each time slot,
users with new service demand could decide either to access or sense one of the BSs within its range.
We denote the action as \(\Phi = \Phi_{a}^i\) for accessing the \(i^{th}\) BS,
and \(\Phi = \Phi_{s}^i\) for sensing.
In the case of sensing,
the BS will respond to the user by sending its next-time-slot system state
in a short message which requires negligible energy consumption.
In the case of accessing, the user sends a request to the chosen BS.
When the request is achievable, the BS activates user's transmission immediately.
Otherwise, the BS declines the request and
informs the user its next-time-slot system state, again with a low energy consuming short message.
In the above process, the short message could
be used as observation by a user to update his/her belief of the BS state (e.g., battery state, number of users in the system).
Therefore, the user's action has to be carefully chosen.
On one hand, the users are intent to maximize their utility by making enough access attempts.
On the other hand, sensing is necessary, as the lack of information will result in useless attempts, causing energy waste and access failures.\\
\indent
From the BSs' perspective, certain protections are needed,
as malicious users could keep accessing one BS and use up all the energy.
To focus our work on formulation from the users' perspective and
not to be distracted by protection details,
a simple protection is used in our work.
If a BS is currently not serving any users and the battery is low,
the BS would reserve the last quantity of energy that could serve a user
for one time slot and forbid new users from using it.
Meanwhile, in our work, we consider rational users who could observe multiple BSs and
chooses an action in every time slot to maximize the number of successful access.
The number of the observed BSs by the user is denoted as \(N_A\).

\section{POMDP BS Access Strategy}
As POMDP could solve decision making problems of different decision horizon lengths under uncertainty,
it perfectly depicts the above EH Super Wi-Fi model.
In order to formulate the problem, two key components in the system model, user number and battery, are carefully considered.
\subsection{User Model}
As in traditional wireless network, the users arrive and leave the BS with certain probability \cite{5}.
Between adjacent time slots, new users may arrive and
old users may leave or be forced to leave when battery is insufficient.
A revised birth and death process that considers forced leaving are proposed as follow,
\begin{align}\label{formula1}
&\zeta\left(S_U'| S_U, Q_B, \Phi\right) = \nonumber\\
&\begin{cases}
	\lambda, &\mbox{if $Q_B$ is enough and $S_U' = S_U + 1$,}\\
	\mu S_U, &\mbox{if $Q_B$ is enough and $S_U' = S_U - 1$,}\\
	I_0\left(S_U'\right), &\mbox{if $Q_B$ is insufficient,}\\
	0, &\mbox{otherwise.}\\
\end{cases}
\end{align}
In the above equation, \(\lambda\) is the arriving rate and
the \(\mu' = \mu S_U\) is the leaving rate of users.
The forced leaving is formulated by the indicator function \(I_0\left(S_U'\right)\),
which indicates that \(S_U'\) next time slot is sure to be \(0\).
Energy depletion happens when \(Q_B- E_T \leq 0\).
\subsection{Battery Formulation}
The harvested energy \(E_T\) is determined by the environmental parameters.
Gaussian models have been proven effective in predicting solar intensity \cite{gaussian,data}.
The harvest model assumes that the solar intensity in a long time period are Gaussian distributed, and the solar intensity during one single time slot, the length of which denoted as \(T_L\), remains unchanged.
Thus, the solar intensity \(W_e\) could be formulated as
Gaussian distributed \(\mathcal{N}\left(x;\mu_S,\sigma_S\right)\)
with average intensity \(\mu_S\) and variance \(\sigma_S^2\).
In current devices,
the harvesting power per reference solar intensity is
\(E_H = W_eJ_{op}V_{op}\Omega_ST_L\eta\), where the \(J_{op}\) and \(V_{op}\) is the optimal operating point,
the \(\Omega_S\) is the number of solar cells and the \(\eta\) is the efficiency \cite{physic}.\\
\indent The transmitting power is set by BSs to provide enough SINR for the receiving users.
Current power adjustment algorithms that use feedback are not valid,
as the system state are fast changing between time slots.
Therefore here a static power management is implemented,
where the BS sacrifices some energy to insure a successful transmission every time slot.
The power consumption in a BS is determined by the number of accessing users,
which is denoted as \(E_T = \Upsilon_T(S_U, Q_B, \Phi)\).
As wide range telecommunication has negligible between-user interference,
we assume the transmission power is proportional to the number of serving users, i.e.,
\(\Upsilon_T(S_U, Q_B, \Phi) = P_T(S_U + \Theta(S_U,Q_B,\Phi_a))\), where \(P_T\) is the transmission power for single user.
\(\Theta(S_U,Q_B,\Phi_a)\) is \(1\) if the rational user could successfully access, otherwise it is \(0\).
Correspondingly, we assume the battery volume \(B_M = \rho P_TT_L\), where \(\rho\) is an integer.
Note that for the nonlinear transmission power function,
our method could still work by setting different battery levels in the below sections.
When the required battery is more than the BS's remaining battery,
the BS will provide a best effort service.
Users with the higher priority are served.
In our case, the rational user has the lowest priority.
Given the \(E_H,\,E_T\), the battery in the next time slot could be calculated as,
\begin{equation}
	Q_B' = \mbox{min}\{Q_B + E_H - E_T, B_M\}.
\end{equation}
\indent In EH powered BSs, the battery is a continuous value.
But in a POMDP, the states have to be discrete.
Intuitively, we could use more battery states to approximate continuous value,
but this brings much increase in the complexity of the algorithm.
Luckily, a certain number of discrete levels could provide enough information during the decision making.
We could set battery levels according to the possible energy consumption in each time slot \(T_L\Upsilon_T(S_U,Q_B,\Phi)\).
In the case of linear power function, the levels are set as \(S_B = \lfloor Q_B / \left(P_TT_L\right) \rfloor\).
And the number of battery states is \(N_B = B_M / \left(P_TT_L\right) +1 = \rho + 1\).
Thus, by knowing the battery states,
the user would know whether the battery is sufficient for transmission.\\
\indent
The prediction of future battery states are based on transition probability between battery states.
In order to calculate the transition probability,
we assume the fluctuate of discrete battery state is quasi-static.
i.e., the residue energy \(Q_B - S_BP_TT_L\) is uniformly distributed between \(\left[0,\,P_TT_L\right)\).
Although errors are brought by this assumption, the quasi-static assumption is proven to be effective after a large number of time slots \cite{data}.
We denote the change of battery state between time slot as \(\Delta_B = S_{B}' - S_B\).
Event \(\xi_j\) represents that the real battery quantity change \(E_H - E_T\)
is equivalent to more than \(j\) but less than \(j+1\) battery state change, namely
\(\xi_j := \{j\leq \frac{E_H - E_T}{P_TT_L} \le j+1\}\).
Then the probability that the battery state will change by \(\Delta_B\)
given \(E_H\) and user action can be computed as follows
\begin{align}&\mbox{Pr}\left(\Delta_B = i |\Phi, E_H, \xi_j \right)\nonumber\\
=&\begin{cases} \frac{\left(E_H - E_T\right)}{P_TT_L} -j, &\mbox{$i = j + 1$},\\
\left(j+1\right) -\frac{\left(E_H - E_T\right)} {P_TT_L}, &\mbox{$i = j$},\\
0, &\mbox{otherwise.}\\
\end{cases}
\end{align}
In the equation, as mentioned in previous section, \(E_T = \Upsilon_T(S_U, Q_B, \Phi)\).
When \(E_H - E_T \le 0\), the probability could be calculated the same way.
The battery transition can be written as
\begin{equation}\label{battery}
\begin{aligned}
	&\mbox{Pr}\left(\Delta_B = i |\Phi\right) = \\
	&\int\nolimits_{i\epsilon_T + E_T}^{\left(i+1\right)\epsilon_T+ E_T}
	\mbox{Pr}\left(\Delta_B = i |\Phi, E_H, \xi_i\right) \mathcal{N}\left(E_H;\bar{\mu_S},\bar{\sigma_S}\right) dE_H+\\
	& \int_{\left(i-1\right)\epsilon_T+ E_T}^{i\epsilon_T + E_T}
	\mbox{Pr}\left(\Delta_B = i |\Phi, E_H, \xi_{i-1}\right) \mathcal{N}\left(E_H;\bar{\mu_S},\bar{\sigma_S}\right) dE_H.\\
\end{aligned}
\end{equation}
In the equation, \(\epsilon_T = P_TT_L\) and \(\bar{\mu_S},\,\bar{\sigma_S}\)
are scaled from \(\mu_S,\,\sigma_S\) after the multiplication with harvesting device coefficients.
\subsection{System Transition Probability}
The POMDP state is the overall system state, which combines all the BSs' system state.
To make the following math more readable and flexible,
the system state \(S\) and \(S_D = \{S_B^1,\,S_U^1,\,\ldots,\,S_B^{N_A},\,S_U^{N_A}\}\)
are equivalent and used simultaneously.
We have \(S = 1,\,2,\, \ldots\,N_S\), where \(N_S = \left(N_BN_U\right)^{N_A}\).
We first calculate the transition probability for a single BS.
From conditionally independence we have
\begin{equation}
\begin{aligned}
	\mbox{P}\left(S_U',S_B'|S_U,S_B,\Phi\right) =
	\zeta\left(S_U'|S_U, S_B, \Phi\right) \delta\left(S_B'|S_U, S_B, \Phi\right).\\
\end{aligned}
\end{equation}
The \(\zeta\left(S_U'|S_U, S_B, \Phi\right)\) is given in equation \eqref{formula1}.
And the battery transition is calculated based on the equation \eqref{battery} as follows
\begin{align}
	&\delta\left(S_B'|S_U, S_B, \Phi\right)\nonumber\\
	= &
	\begin{cases}
		\mbox{Pr}\left(\Delta_B = S_B' - S_B|\Phi \right), &\mbox{if $S_B' \le N_B - 1$,}\\
		\sum_{\Delta_B = S_B' - S_B}^{\Delta_B = \Delta_B^{Max}}\mbox{Pr}\left(\Delta_B|\Phi\right),
		&\mbox{if $S_B' = N_B - 1$.}\\
\end{cases}
\end{align}
When the battery is fully charged, \(S_B'=N_B - 1\), all the extra harvested battery is abandoned.
Note that we truncate the probability for \(\Delta_B >\Delta_B^{Max}\) as they are as small as zero by \(4\) decimals.
Thus the POMDP state transition is computed as
\begin{align}\label{transition}
	T\left(S'|S,\Phi\right) = \prod_{i = 1}^{i = N_A}\mbox{P}\left(S_B^{i,'}, S_U^{i,'}|S_U^i, S_B^i, \Phi\right).
\end{align}
\subsection{Observation Function and POMDP Iteration Algorithm}
In POMDP formulation, the user only has the partial knowledge of the system.
As mentioned above, the user could get the target BS's next-time-slot system state
\(S_B^O\) and \(S_U^O\) at the end of each time slot.
We use observation \(O\) to represent \(S_B^O,\,S_U^O\).
The observation probability function given the system state in the next time slot is
\begin{align}
	Z\left(O|S',\Phi\right) = \mbox{Pr}\left(O\Big|S_U^{t,'}, S_B^{t,'}\right) =
	I_{S_U^{t,'},S_B^{t,'}}\left(S_U^O, S_B^O\right).
\end{align}
\(S_B^{t,'}\) and \(S_U^{t,'}\) are the system state of the target BS next time slot.
The indicator function is \(1\) when \(S_U^{t,'} = S_U^O,\,S_B^{t,'}=S_B^O\) or \(0\) otherwise.\\\indent
The reward is define as \(R = 1\) if the access succeeds, else \(R= 0\).
Then the value function of a single state is
\begin{equation}
\begin{aligned}
	V_t^\Phi\left(S\right) = R\left(S,\Phi\right) +\gamma\sum\limits_{S'}T\left(S'|S,\Phi\right)V_{t-1}^\pi\left(S'\right),
\end{aligned}
\end{equation}
where \(\pi\) denotes the optimal action in that state.
As no full knowledge is held for the user, we use a belief vector to denote the user's system state belief
\(\beta = \lbrack \beta\left(S = 1\right),\,\beta\left(S = 2\right),\,\ldots,\,\beta\left(S = N_S\right)\rbrack\).	
Then the particular value function with a certain belief \(\beta\) is given by
\begin{equation}
\begin{aligned}
	V_t^\Phi\left(\beta\right) = & \sum\limits_{S}R\left(S,\Phi\right)\beta\left(S\right) +\\
	&	\gamma\sum\limits_{S}\sum\limits_{S'}\beta\left(S\right)T\left(S'|S,\Phi\right)V_{t-1}^\pi\left(S'\right).
\end{aligned}
\end{equation}
For simplicity, if we already know all the value function of at the time \(t-1\) during iterations,
an alpha value vector \(\alpha_t^\Phi = \lbrack \alpha_t^\Phi\left(S = 1\right),\,
\alpha_t^\Phi\left(S = 2\right),\,\ldots,\,\alpha_t^\Phi\left(S = N_S\right)\rbrack\)
could be used to simplify the value function as
\(V_t^\Phi\left(\beta\right) = \sum\beta\left(S\right)\alpha_t^\Phi\left(S\right)\).
Thus, the optimal action can be given by
\begin{equation}
\begin{aligned}
	\pi\left(\beta\right) =
	\arg\underset{\alpha_t^\Phi}{\max}\sum\limits_{S}\beta\left(S\right)\alpha_t^\Phi.\\
\end{aligned}
\end{equation}
The corresponding value function \(V_t^\Phi\left(\beta\right)\) could be calculated using action \(\pi\left(\beta\right)\).
However, the corresponding optimal policy is not as easy as it seems to be, as the \(\beta\) has a continuous value,
and even for the same \(\Phi\) and \(t\), there are still multiple possible \(\alpha\) vectors during iterations.
But fortunately, the \(V_t\) could be regarded as the function value of \(\beta\) in a hyper coordinate system,
the axes of which are the components of \(\beta\).
As each set of \(\alpha_t\) vector could be regarded as a set of parameters of a hyper linear function,
there is a dominated hyperplane structure in the model.
The continuous belief space is divided by several \(\alpha\)-vector-dominated hyperplanes into several partitions.
The partitions of belief space in time \(t\) could be calculated given all the dominating \(\alpha_{t-1}\).
The details of algorithm for solving the partitions could be find in a well written tutorial \cite{pomdptool}.\\
\indent
After obtaining the optimal action, the user could act accordingly,
and update its belief vector after receiving the observation by using the following formula.
\begin{align}
	\beta'\left(S'|\Phi, O\right) = \frac{\sum_{O}Z\left(O|S',\Phi\right)\sum_{S}
	T\left(S'|S,\Phi\right)\beta\left(S\right)}
	{\sum_{S'}\sum_{O}Z\left(O|S',\Phi\right)\sum_{S}T\left(S'|S,\Phi\right)\beta\left(S\right)}.
\end{align}

\subsection{Suboptimal Access Policy}
The optimal POMDP solution could be calculated off-line within seconds when the number of states is small.
However, the use of POMDP method is limited when \(N_A\) is massive,
and when the environment parameters, like solar coefficients \(\mu_S\), \(\sigma\),
the birth rate \(\lambda\) and death rate \(\mu\) of users, change quickly.\\
The POMDP formulation will maximize success access ratio \(\eta_A = N_S/N_T\),
where \(N_S\) is number of success access and \(N_T\) is the number of the time slots.
We propose a dual perspective of solving the problem by focusing on the harvested energy.
We name it Energy Based (EB) Method.
The problem is reformulated as,
\begin{equation}
\begin{aligned}
	\underset{\Phi_t,t=0,\,\ldots,\,N_T-1}{\max}\sum\nolimits_{t=1}^{N_T}
	\mbox{E}\big[\,\mbox{H}\left(\beta^t, \Phi_t\right)\big],\\
\end{aligned}
\end{equation}
where the expectation function E[\(\cdot\)] considers the probability of receiving different observations
and thus having different belief vector \(\beta^t\), and $\mbox{H}\left(\beta^t, \Phi_t\right)$ is the overall harvesting energy in all the BSs as follows
\begin{equation}
\begin{aligned}
	\mbox{H}\left(\beta^t, \Phi_T\right) = \sum\beta^t\left(S\right)\sum_{i=1}^{N_A}
	\mbox{min}\left(E_H^i,\,E_T^i+B_M-Q_B^i\right)
\end{aligned}
\end{equation}
In such a case, a suboptimal could be proposed by maximizing the system's next-time-slot harvested energy, i.e.,
\begin{equation}
\begin{aligned}
	\Phi\left(\beta^t\right) = \arg{\max}\,\,\mbox{E}\big\lbrack\,\mbox{H}(\beta^{t+1}, \Phi_t)\big\rbrack,\\
\end{aligned}
\end{equation}
When BS sensing is the optimal action, the user will choose to sense the BS that is not sensed for the longest time.
Due to the limited space, some key rationality of the EB method is summarized.
First of all, when the solar intensity is strong,
we could assume few users will be forced to leave and thus the consumed energy by them are stable.
As the transmission power is proportional to number of users,
we could deduce that the more energy the system harvests,
the more utility the rational user could achieve.
Besides, the EB method is an unselfish method, which would sacrifice some reward,
but tends to protect overall utility.
And EB method could use learning algorithm to adjust to quick environmental changes.
\section{Simulation Results}
\begin{figure}
\centering
\includegraphics[width=0.5\textwidth]{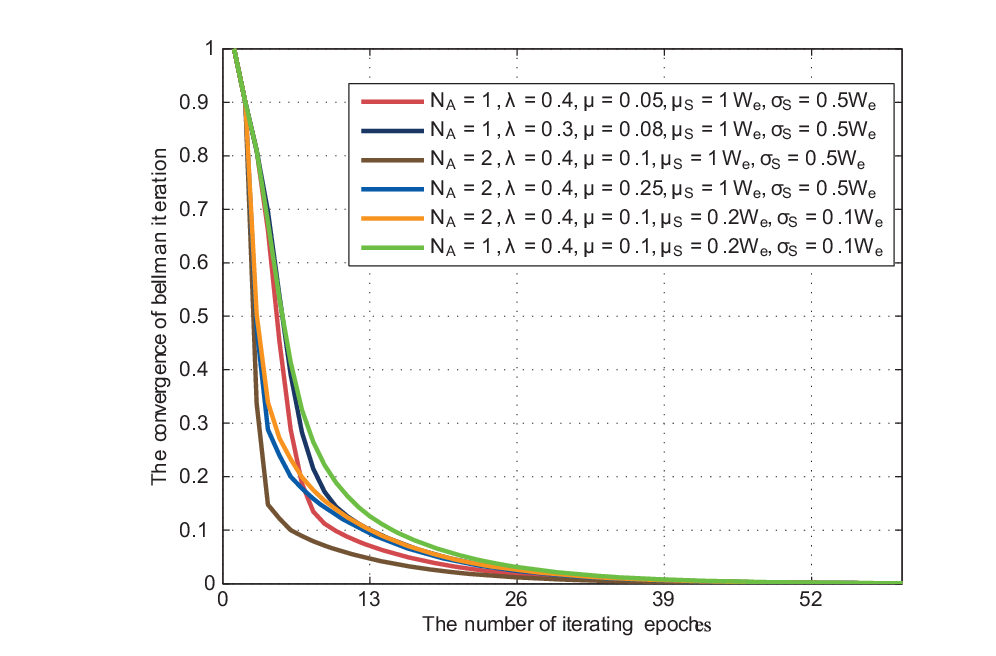}
\caption{Illustration of the convergence of the POMDP iteration algorithm}
\end{figure}
\begin{figure}[t]
\centering
\subfigure[The utility ratio with arrival rate in single BS]{
\includegraphics[width=0.5\textwidth,height=6.7cm]{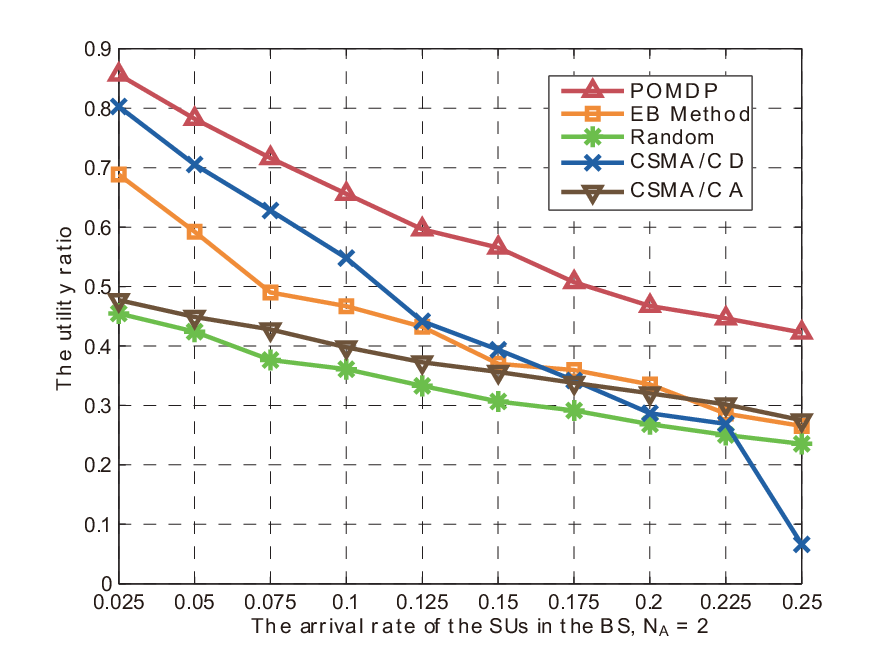}}
\subfigure[The utility ratio with solar intensity in single BS]{
\includegraphics[width=0.5\textwidth,height=6.7cm]{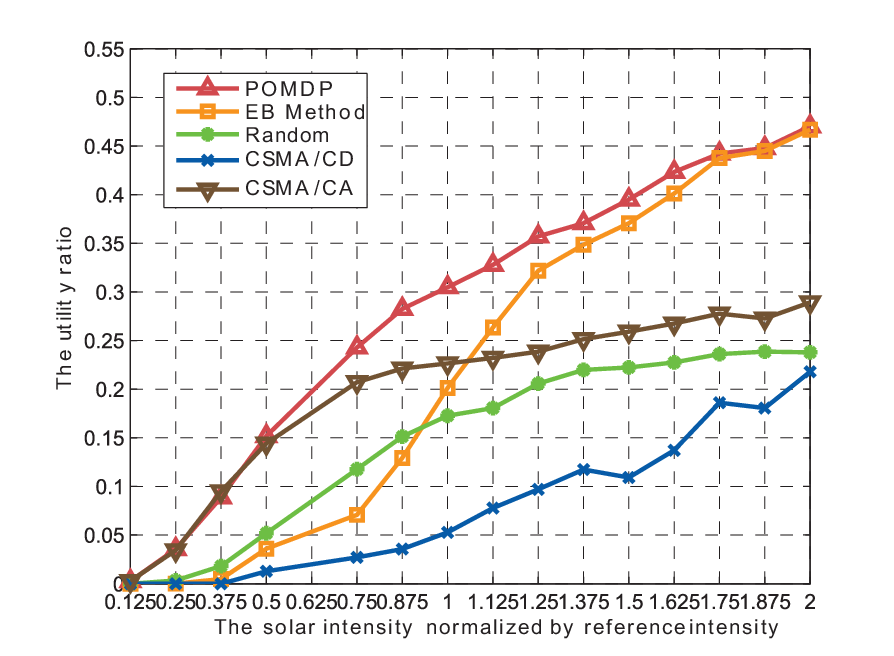}}
\caption{Single BS with \(S_U = 0,\,1,\,2,\,3,\, N_B = 8\)}
\end{figure}
\begin{figure}[t]
\centering
\subfigure[The utility ratio with arrival rate in two BS]{
\includegraphics[width=0.5\textwidth,height=6.7cm]{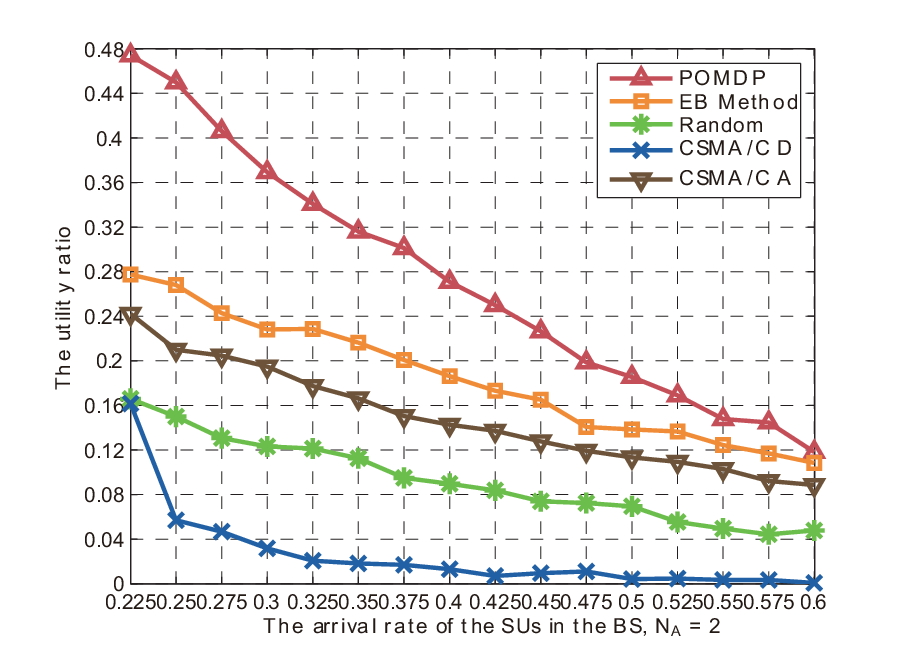}}
\subfigure[The utility ratio with solar intensity in two BS]{
\includegraphics[width=0.5\textwidth,height=6.7cm]{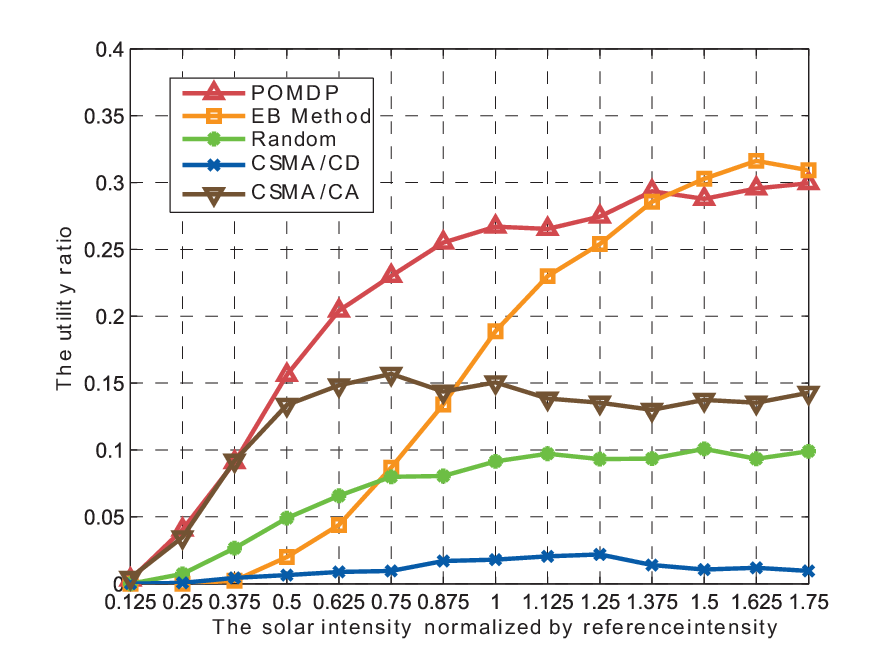}}
\caption{Two BS with \(U_N = 0,\,1,\, N_B = 3\)}
\end{figure}
In this section, the convergence and effectiveness of the algorithm are illustrated.
Fig. 2 shows the convergence of the POMDP iteration algorithm, where the y-axis means a user's value difference between two adjacent iterations and the discount factor is \(\gamma = 0.9\).
A Bellman stopping criteria is used to determine the stop of iteration.
In the figure, the \(\alpha\) vector error shows the convergence of the iteration algorithm.
The POMDP simulation tool is provided by \cite{pomdptool}.
As we can see from the results, our proposed POMDP algorithm converges exponentially under various parameter settings.\\
\indent The effectiveness of the algorithms is validated by comparing the \(\eta_A\)
under the overall time slots \(N_T = 10000\).
In the simulation, parameters are used as follow.
The reference benchmark solar intensity is given as \(\mathcal{W}_e = 1\mbox{kW}/m^2\),
which is the average intensity on the surface of Earth \cite{electric}.
We use the data from work \cite{circuit}, where the optimal power per benchmark solar intensity \(\mathcal{W}_e\)
is \(P_H = J_{op}V_{op} = 1.32\mbox{mW}/\mathcal{W}_e\), with a efficiency \(\eta = 75 \%\).
And there are \(\Omega_S = 40\) cells in one harvesting device.
The time slot length is \(T_L = 200\mbox{ms}\).
Transmission power for serving one user is \(P_T = 40\mbox{mW}\).\\
\indent In order to show the efficiency, several algorithms are implemented as comparisons.
including Carrier Sensing Multiple Access/Collision Avoidance, Collision Detection (CSMA/CA and CSMA/CD).
Note that here the ``carrier sensing'' means that one user senses the BSs, i.e., receives the short message from one BS, instead of the physical carrier.
The CSMA/CD method stops new request when detecting a failure, and then an exponential back-off algorithm is used.
After \(c\) failures, a random number of sleeping time slot between \(0\) and \(2^c - 1\) is chosen.
The CSMA/CD method would access the BS after the user senses the BS in the last time slot
and knew that a successful service is available.
In random access algorithm, the user simply chooses action with equal probability.\\
\indent In Fig. 3, we consider single BS with possible number of users from \(0\) to \(3\), \(N_B = 8\),
and the leaving rate \(\mu = 0.05\).
In Fig. 3-(a), the \(\mu_S = 1\) and \(\sigma_S = 0.5\), while in Fig. 3-(b), the arrival rate \(\lambda = 0.4\).
As shown in the figure, the performance of proposed POMDP is significantly higher,
with the EM method's overall performance following at the second place,
which validates the efficiency of our algorithms.
From Fig. 3-(a), we can see that the CSMA/CD has a good performance when the system is not busy,
but the performance deteriorates quickly with the increasing arrival rate of users.
In Fig. 3-(b), one prominent point is that even when the solar intensity is strong,
the utilities of the traditional algorithms' are saturated, failing to further enhance users' utility, due to the reason that those algorithms are not able to make use of the EH information of the system.
It is also worth mentioning that, as we predicted, when the solar intensity is strong,
the suboptimal EB method will approach the proposed POMDP method, which can decrease the complexity to a large extent.\\
\indent In Fig. 4, multiple BSs are considered, with possible number of user from \(0\) to \(1\),
\(N_B = 3\), and the leaving rate \(\mu = 0.05\).
In the figure, we find that when the possible serving positions is limited, the crowded system makes the CSMA/CD method almost useless.
In (b), a simple analysis of the CSMA/CD performance could be given.
When the solar intensity is small, the utility of CSMA/CD increases with intensity due to more available sources.
But when the solar intensity is strong, as the more users are staying in the BS,
the user has less chances of being served, and the utility decreases.
Also, in (b), the suboptimal EB method could outperform POMDP method when the intensity is strong,
which is mainly brought by the approximation in the POMDP formulation, such as using the quasi-static approximation.
\section{Conclusion}
In this paper, we proposed a powerful POMDP algorithm to solve the access problem in EH powered network,
which is promising and instructive in building a national range Super Wi-Fi network.
The framework given in this paper is adjustable to EH problems other than the Solar EH one.
To reduce complexity and adjust to environmental changes, a suboptimal EB method is proposed as well.
The effect of solar intensity, user arriving rate, leaving rate and many other features are considered,
proving our work reliable and effective.
Future work of this paper could focus on the prediction of system parameters and multiuser accessing scenario.
\bibliographystyle{IEEEtran}
\bibliography{Ref}
\end{document}